\documentclass[onecolumn]{elsart3p}
\journal{Physica D}
\usepackage{lineno}
\usepackage{amsfonts}
\usepackage{amssymb}
\usepackage{amsmath}
\usepackage{epsfig}
\usepackage{graphicx}
\usepackage{color}

\begin{document}

\begin{frontmatter}

\title{Variational approximations in discrete nonlinear Schr\"{o}dinger
equations with next-nearest-neighbor couplings}

\author[chong]{C. Chong\corauthref{cor1}}
\corauth[cor1]{Corresponding author},
\ead[url]{http://www.iadm.uni-stuttgart.de/LstAnaMod/Chong/home.php}
\author[sdsu]{R.\ Carretero-Gonz\'{a}lez},
\author[bam]{B.A.\ Malomed}, and
\author[pgk]{P.G.\ Kevrekidis}

\address[chong]{%
Institut f\"ur Analysis, Dynamik und Modellierung,
Universit\"at Stuttgart, Stuttgart 70178, Germany
}
\address[sdsu]{%
Nonlinear Dynamical Systems Group{$^1$},
Computational Sciences Research Center, and\\
Department of Mathematics and Statistics,
San Diego State University, San Diego,
CA 92182-7720, USA
}
\thanks[nlds]{{\tt URL:} http://nlds.sdsu.edu/}
\address[bam]{%
Department of Physical Electronics, School of Electrical Engineering,
Faculty of Engineering, Tel Aviv University, Tel Aviv 69978, Israel
}
\address[pgk]{%
Department of Mathematics and Statistics, University of
Massachusetts, Amherst MA 01003-4515, USA
}

\begin{abstract}
Solitons of a discrete nonlinear Schr\"{o}dinger equation which
includes the next-nearest-neighbor interactions are studied by
means of a variational approximation and
numerical computations. A large family of multi-humped solutions,
including those with a nontrivial phase structure which are a
feature particular to the next-nearest-neighbor interaction model, 
are accurately
predicted by the variational approximation. 
Bifurcations linking solutions with the trivial
and nontrivial phase structures are also captured remarkably well,
including a prediction of critical parameter values.
\end{abstract}

\maketitle



\end{frontmatter}

\section{Introduction}

\label{sec:model}

It has long been known that the discrete nonlinear Schr\"{o}dinger
(DNLS) equation is a relevant model for a wide range of applications
including nonlinear optics (waveguide arrays), matter waves
(Bose-Einstein condensates trapped in optical lattices) and
molecular biology (modeling the DNA double strand). One of the
reasons this model and its variants are relevant in many areas is
the extensive range of phenomenology that the equations encompass,
including the discrete diffraction, gap solitons, Peierls-Nabarro
potentials, lattice chaos, Anderson localization, snaking and
modulation instabilities, among other effects, see the recent
book~\cite{PK09} for a review.

In this work we consider the DNLS equation which allows for linear coupling
to additional (than just nearest) neighbors. Such ``nonlocal'' 
interactions have been studied before,
and, in particular, solutions with a nontrivial phase distribution \cite%
{PKnnn} 
were identified, while 
a sufficiently slow decay of the interaction
strength was found to
lead to bistability of the fundamental soliton solutions \cite%
{Gaididel97,Rasmussen98}. As concerns physical realizations of such
interactions, they are relevant in models of waveguide arrays that are
aligned in a zigzag pattern \cite{zigzag02}, and in modeling the charge
transport in biological molecules \cite{Mingaleev02}. For other applications
of the nonlocal DNLS systems see Ref. ~\cite{PKnnn} and references therein.

The nonlocal DNLS equation has the general form
\begin{equation}
i\dot{u}_{n}+|u_{n}|^{2}u_{n}=-\epsilon \sum_{m\in {\mathbb{N}}}k_{nm}u_{m},
\label{eq:dnls}
\end{equation}%
where $u_{n}(t)$ is the complex discrete wave field, $n$ is the integer
lattice coordinate, and the real 
parameter $\epsilon $ is the coupling strength, the
coupling matrix being composed of real symmetric
elements $k_{nm}$. The Hamiltonian and
power,
\begin{equation}
\mathcal{H}=\sum_{n\in {\mathbb{N}}}\left[ \frac{1}{2}|u_{n}|^{4}+\epsilon
\sum_{m\in {\mathbb{N}}}k_{nm}u_{n}^{\ast }u_{m}\right] ,~\mathcal{P}=\Vert
u\Vert _{l^{2}}^{2},
\end{equation}%
where $\Vert u(t,\cdot )\Vert _{l^{2}}^{2}=\sum_{n\in {\mathbb{Z}}%
}|u_{n}(t)|^{2}$, are conserved quantities of Eq.~\eqref{eq:dnls} if the
coupling matrix is symmetric. Solutions of Eq.~\eqref{eq:dnls} are obviously
invariant against the phase shift, $u_{n}(t)\leftrightarrow u_{n}(t)\exp (%
\mathrm{i}\beta )$ with real $\beta \in \mathbb{R}$ (the gauge invariance),
and against the reflection transformation, $u_{n}(t)\leftrightarrow
u_{-n}(t) $. The case of $k_{nm}=\delta _{m,n\pm 1}-2\delta _{m,n}$, where $%
\delta _{m,n}$ is the Kronecker's delta-symbol, corresponds to 
nearest-neighbor interactions (discrete Laplacian) 
of the standard DNLS equation. The goal
of this work is to develop a variational approximation (VA) for localized
solutions (discrete solitons) of Eq.~\eqref{eq:dnls} with the extended
linear coupling. Predictions provided by the VA will be verified by
comparison with numerical results.
The manuscript is organized as follows.
In Sec.~\ref{SEC:ss} we present the equations to be solved for
the computations of the steady states, 
their phase condition for existence from the anti-continuous limit, 
their dynamical reduction to a four-dimensional map, 
and the stability and bifurcations for the basic type
of solutions that we will consider.
In Sec.~\ref{SEC:VA} we describe in detail the VA employed
to describe the main type of discrete solitons supported
by the next-nearest-neighbor coupling that include trivial
phase (all nodes with same or opposite phases) 
and nontrivial phase configurations.
Finally, in Sec.~\ref{SEC:conc} we present our conclusions and
potential avenues for future work.

\begin{table}[t]
\begin{center}
\begin{tabular}{|c|c|c|c|c|c|c|}
\hline
~$n$~ & ~~-2~~ & ~~-1~~ & ~~0~~ & ~~1~~ & ~~2~~ & ~label~ \\ 
\hline
~ & - & - & 0 & - & - & a \\
~ & - & - & 0 & 0 & - & b \\
~$\theta_n$~ & - & 0 & - & 0 & - & c \\
~ & - & 0 & 0 & 0 & - & d \\
~ & - & 0 & $\pi$ & 0 & - & e \\
~ & - & $-.58$ & $\pi$ & $.58$ & - & f \\ \hline
\end{tabular}
\label{tab:profiles}
\end{center}
\caption{Examples of trivial (a)--(e) and nontrivial (f) phase
distributions referred to throughout the text. Dashes represent
non-excited sites, hence they do not carry phase. The corresponding
solutions, which have the form of $\protect\phi _{n}=\exp
(\mathrm{i}\protect\theta _{n})$ in the
anti-continuum limit, persist for $\protect\epsilon \neq 0$. Parameters are $%
k_{2}=0.6,k_{1}=1.0$ and $k_{0}=-2(k_{2}+k_{1})$. The $n=\pm 1$ phase values
listed for (f) are approximate ones. Their exact values are given by 
Eq.~\eqref{eq:cond}. }
\end{table}

\section{Steady states}\label{SEC:ss}

\label{sec:steady}

A natural starting point is to consider steady-state solutions, in the form
of
\begin{equation}
u_{n}(t)=\phi _{n}e^{\mathrm{i}t}  \label{eq:steady}
\end{equation}%
where amplitudes $\phi _{n}$ may be complex, and rescaling was used to fix
the frequency as $-1$. Upon the substitution of expression
\eqref{eq:steady} into Eq.~\eqref{eq:dnls}, we arrive at the stationary
problem
\begin{equation}
(|\phi _{n}|^{2}-1)\phi _{n}=-\epsilon \sum_{m\in {\mathbb{N}}}k_{nm}\phi
_{m}.  \label{eq:dnls-sta}
\end{equation}%
Throughout the manuscript, we will consider the next-nearest-neighbor (NNN)
coupling as an example to illustrate salient features of the nonlocal model.
In addition, we assume that the coupling matrix is symmetric and the lattice
is uniform, hence the NNN variant of Eq.~(\ref{eq:dnls-sta}) is, with
obviously redefined elements of the matrix:
\begin{equation}
\label{eq:dnls-nnn}
\begin{array}{l}
(|\phi _{n}|^{2}-1)\phi _{n}=\\
\qquad
-\epsilon (k_{2}\phi _{n-2}+k_{1}\phi
_{n-1}+k_{0}\phi _{n}+k_{1}\phi _{n+1}+k_{2}\phi _{n+2}).
\end{array}
\end{equation}

\subsection{The phase condition} 

In the anti-continuum
(AC) limit, which is defined by $\epsilon =0$, solutions are defined by
respective sets of excited sites (with a nonzero field at them), taken as $%
\phi _{n}=e^{\mathrm{i}\theta _{n}}$ where $\theta _{n}$ are arbitrary
phases. It was shown in Ref. \cite{PKnnn} that the solutions initiated at
the AC limit persist, as the inter-site couplings are turned on
($\epsilon\not=0$), if the following conditions on the phases are satisfied:
\begin{equation}
\sum_{n\neq m}k_{nm}\sin (\theta _{n}-\theta _{m})=0.  \label{eq:cond}
\end{equation}
In the NNN case, Eq.~\eqref{eq:cond} reduces to either\\
(i) $\sin (\theta _{n}-\theta _{m})=0,$ for all $n,m\in \{1,2,3\}$ or\\
(ii) $\theta_{-1}-\theta _{0}=\theta _{0}-\theta _{1}$ 
and $\cos (\theta _{-1}-\theta _{0})=-k_{1}/(2k_{2})$. 
In general, we consider the relative phases as trivial
if they are integer multiples of $\pi $, and nontrivial otherwise. In this
sense, case (i) is trivial and (ii) is nontrivial. If a solution is composed
of trivial relative phases, we say the solution is trivial (not to be
mistaken for the zero solution, which we ignore), and likewise for the
nontrivial case.

\subsection{Dynamical reduction} 
Equation~\eqref{eq:dnls-sta} may be rewritten as a system of first-order
difference equations. For the NNN coupling, the fourth order recurrence
relation (\ref{eq:dnls-nnn}), reduces to the following four coupled first
order difference equations:
\begin{eqnarray}
\notag
X_{n+1} &=&
\notag
\frac{\phi _{n}-|\phi _{n}|^2\phi_{n}}{\epsilon}
-\left(Z_{n}
+\frac{k_{1}}{k_{2}}Y_{n}
+\frac{k_{0}}{k_{2}}\phi _{n}
+\frac{k_{1}}{k_{2}}X_{n}\right), \\
\notag
\phi _{n+1} &=&X_{n}, \\
\label{eq:map}
Y_{n+1} &=&\phi _{n}, \\
\notag
Z_{n+1} &=&Y_{n}.%
\end{eqnarray}
Discrete soliton solutions of Eq.~\eqref{eq:dnls-nnn} correspond to
homoclinic orbits of the fixed point at the origin, in terms of map 
\eqref{eq:map}. Decay rates of the solutions are given by $\lambda ^{|n|}$,
where $\lambda $ is the eigenvalue of the Jacobian of system \eqref{eq:map}
evaluated at the origin and corresponding to the stable manifold. A
detailed description of dynamical reductions can be found in 
Ref.~\cite[Chapter 11]{PK09}. For our purposes, the only information needed from this
reduction are the decay rates. The actual computation of the manifolds is a
delicate issue, since the two-dimensional manifolds are embedded into the
four-dimensional space for real $\phi _{n}$, and the eight-dimensional space
for complex $\phi _{n}$. Standard methods for the numerical computation of
the manifolds will fail in general due to the existence of additional, more
dominant, eigen-directions. Although the detailed  computation of
the manifolds falls outside the scope of the present work, it would be
interesting to approximate the manifolds with an appropriate parametrized
cubic polynomial following a method similar to that developed in 
Ref.~\cite{Cuevas08} for the nearest-neighbor (local) DNLS.


\subsection{Linear stability} 

\begin{figure}[tbh]
\centerline{
 \epsfig{file=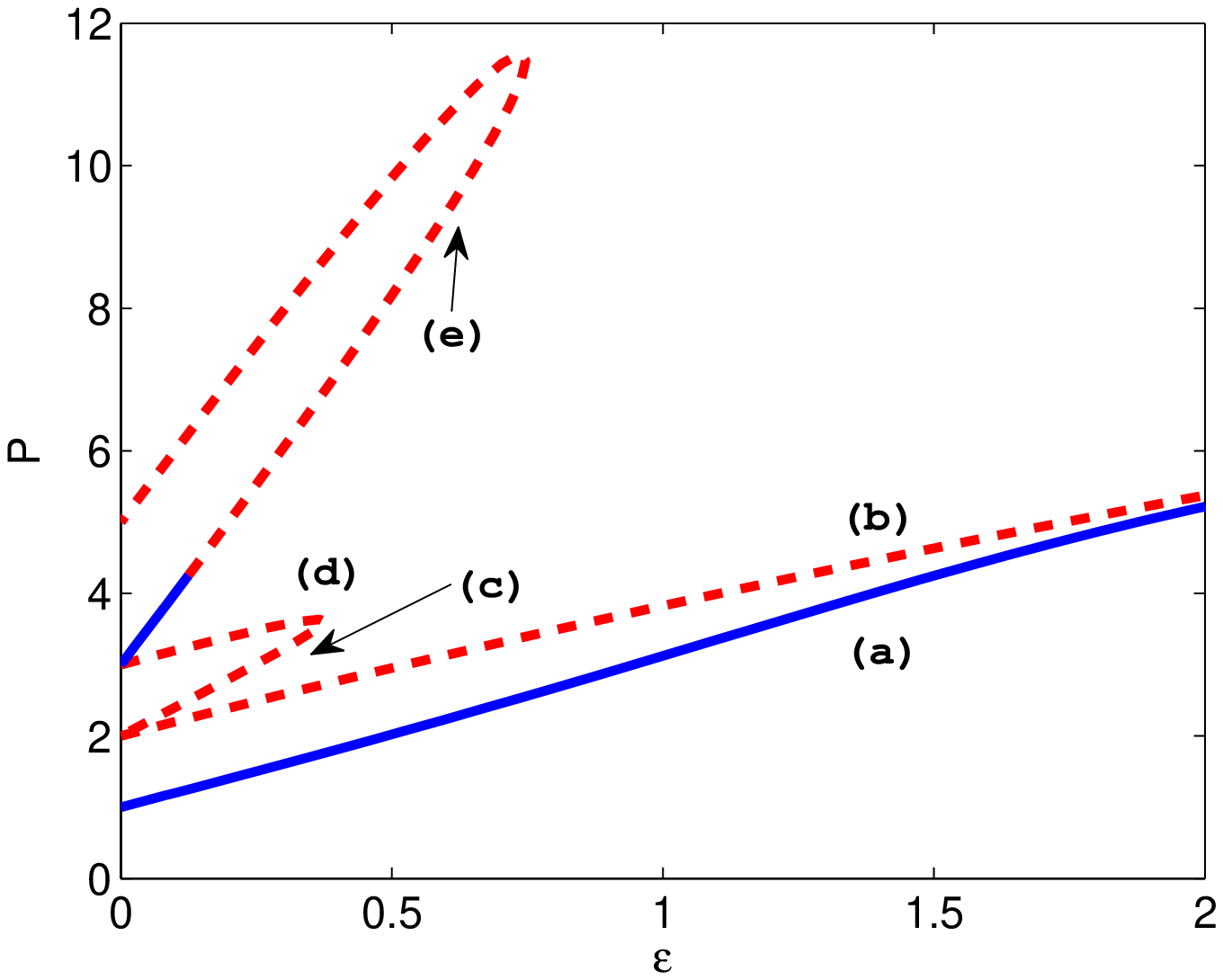,width=8cm,angle=0}
}
\centerline{
 \epsfig{file=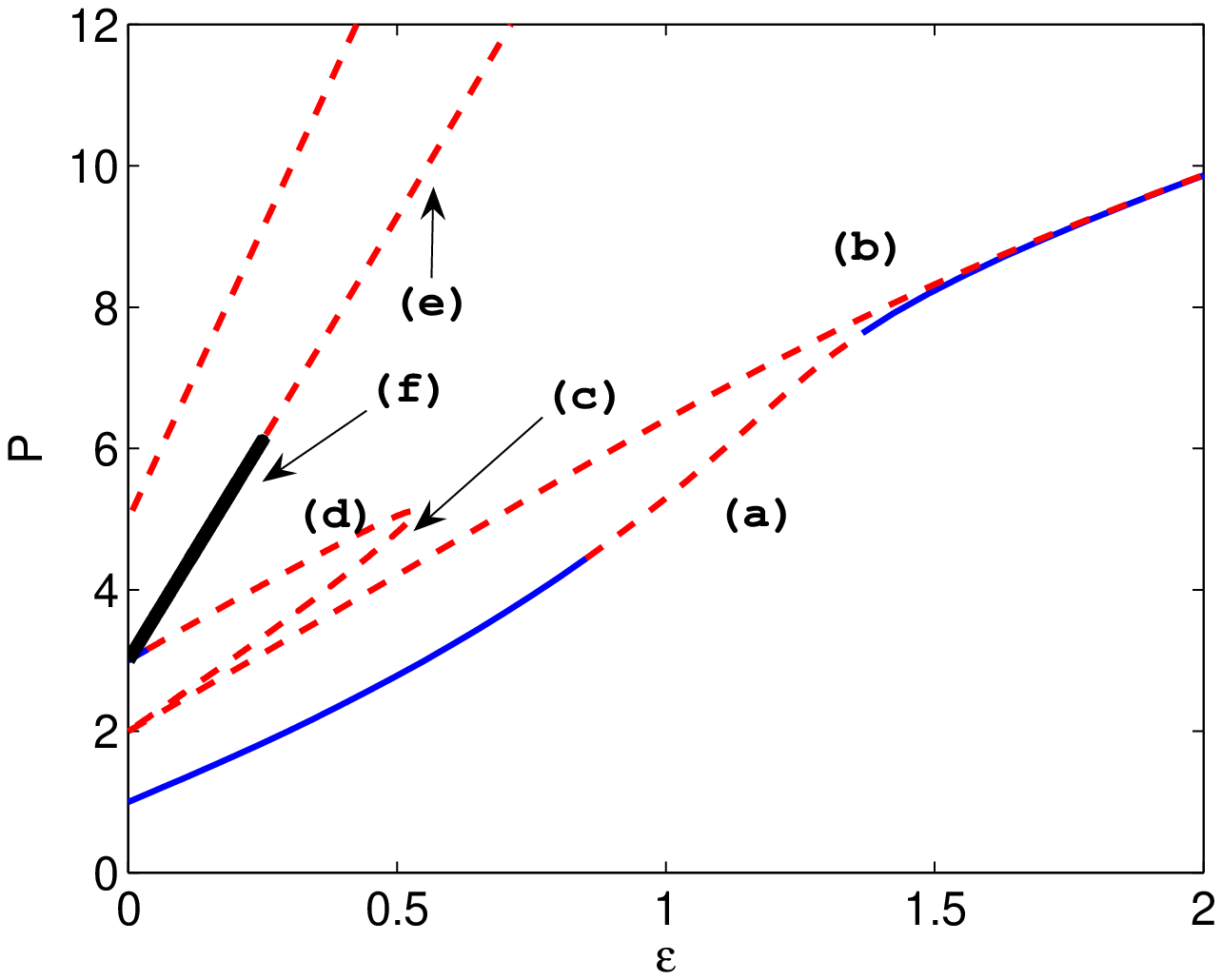,width=8cm,angle=0}
 }
\caption{(Color online) 
Top: The power versus the coupling strength for
solution of types (a)--(e) from Table~\protect\ref{tab:profiles} for the
nearest-neighbor (local) DNLS equation, i.e.~with $k_{2}=0$. 
Dashed red and solid blue lines
correspond to unstable and stable solutions, respectively. 
Bottom: The same
branches in the NNN DNLS equation, with $k_{2}=0.6$. Solution (f) with a
nontrivial phase distribution (black thicker part of the line) appears in
the latter case, with power which is very close to that of solution (e)
with the trivial phase distribution. This bifurcation can be better seen by
comparing the phases, see Fig.~\protect\ref{fig:phases}.
The unlabeled (top) branch (which corresponds to initial phases of
$\{ 0,0, \pi, 0, 0 \}$) merges with the branch labeled
(e) in both panels (although the collision occurs outside the plotting region for
the bottom panel).}
%
\label{fig:bif}
\end{figure}

The linear stability can be
analyzed in the usual way, assuming the perturbed solution as
\begin{equation}
u_{n}(t)=\left( \phi _{n}+(v_{n}+iw_{n})e^{\lambda t}+(v_{n}^{\ast
}+iw_{n}^{\ast })e^{\lambda ^{\ast }t}\right) e^{it},
\end{equation}
which leads to the respective spectral problem,
\begin{equation}
\left\{
\begin{array}{cc}
v_{n}-\epsilon \sum_{m\in {\mathbb{N}}}k_{nm}v_{m}-3|u_{n}|^{2}v_{n} &
=-\lambda w_{n}, \\
w_{n}-\epsilon \sum_{m\in {\mathbb{N}}}k_{nm}w_{m}-|u_{n}|^{2}w_{n} &
=\lambda v_{n}.
\end{array}%
\right. 
\label{eq:eig}
\end{equation}
We are looking for nonzero eigenvectors, i.e., solutions of the linearized
system in the $l^{2}(\mathbb{Z},\mathbb{C}^{2})$ space. The corresponding
steady-state solution is called unstable if there exists at least one
eigenvector for which $\mathrm{Re}(\lambda )>0$. The only solutions that are
stable (for sufficiently weak coupling) 
in the nearest-neighbor DNLS equation are those with consecutive $\pi $ phase
differences between adjacent sites \cite{Todd,PKF}. 
As we explain below, the extended coupling
affects the stability (see Fig.~\ref{fig:bif}).

\subsection{Bifurcations} 

From Eq.~\eqref{eq:cond} we
see that in the standard nearest-neighbor DNLS equation, only solutions with trivial phase
distributions persist for non-zero coupling. These solutions can be
continued in $\epsilon $ and gradually vanish through saddle-node
bifurcations \cite{Konotop}\footnote{Generally, the solutions of the
nearest-neighbor DNLS may disappear through either pitchfork or 
saddle-node (in fact, more appropriately saddle-center) bifurcations.
The configurations considered for this model herein all terminate
in the latter type of bifurcation, as can be confirmed by comparison
with Table 1 of \cite{Konotop}.}, with only the single-site ($\theta _{0}=0$) and
two-site ($\theta _{0}=\theta _{1}=0$) solutions persisting toward the
continuum limit (with both approaching the soliton solution of the
continuous NLS equation). This bifurcation scenario is shown in the top
panel of Fig.~\ref{fig:bif} for configurations (a)--(e) of 
Table~\ref{tab:profiles}. The extended coupling allows for additional types of
solutions with nontrivial phase patterns, which, in the case of the NNN
system, means one additional waveform, indicated by the thick black solid
line in the bottom panel of
Fig.~\ref{fig:bif}. In these diagrams, the power is plotted versus coupling
strength $\epsilon $. For this choice of parameters, the nontrivial branch
(f)  and the trivial 
one (e) lie very close to each other, making the
identification of the bifurcation points difficult. 
The role of the
nontrivial-phase solutions can be better seen through the comparison of the
phases, see below Figs.~\ref{fig:phases} and~\ref{fig:compare_nontriv}.
Also, the extended coupling modifies the stability properties
of some trivial phase solutions. For example, the
branch (a) is always stable for the nearest-neighbor interactions
(see top panel of Fig.~\ref{fig:bif}) while it displays an
instability window for intermediate coupling strengths in the
NNN coupling case (see bottom panel of Fig.~\ref{fig:bif}), in a way
somewhat reminiscent of the findings of \cite%
{Gaididel97,Rasmussen98}.

\section{The Variational Approximation}\label{SEC:VA}

The Lagrangian of Eq.~\eqref{eq:dnls-sta} is
\begin{equation}
\mathcal{L}=\sum_{n\in {\mathbb{Z}}}\left[ \frac{1}{2}|\phi _{n}|^{4}-|\phi
_{n}|^{2}+\epsilon \sum_{m\in S}k_{nm}\phi _{n}^{\ast }\phi _{m}\right] .
\label{eq:Lag-sta}
\end{equation}
According to the variational principle, critical points of the Lagrangian
\eqref{eq:Lag-sta} correspond to solutions of
Eq.~\eqref{eq:dnls-sta}. This underlies the heuristic
justification of the variational approximation (VA), whereby an
ansatz (trial configuration of the wave field) with a finite number
of parameters is substituted into the Lagrangian, and critical
points are then sought for the resulting finite parameter subspace. This
approach has long been used in various applications, 
see the review~\cite{Ma02} and more recent works, such as
Refs.~\cite{semi,Ka05,CP09}. The VA has also been used to study nonlocal
interactions, see, e.g., Ref.~\cite{Rasmussen98}, but in the latter
case only solutions initiated by a single excited site in the AC
limit, i.e., solutions of type (a) in terms of Table 1, were studied.
To the best of our knowledge, the present work for the first 
time extends the VA to describe not only discrete multi-humped solutions with
arbitrary phases (with at least three excited sites), but also ones such
with nontrivial phase distributions.

\subsection{Trivial Phase Distributions}

We start by considering solutions with nontrivial phase
distributions. Our objective is to construct approximate discrete solitons
by means of a real-valued ansatz:

\begin{equation}
\begin{array}{ll}
\psi _{n} & =\left\{
\begin{array}{ll}
B_{n}~\mathrm{for} & n\in S, \\
A\exp (-\eta |n+n_{0}|)~\mathrm{for} & n\notin S,%
\end{array}%
\right.  \\[2ex]
&
\end{array}
\label{eq:ansatz}
\end{equation}
where $S$ denotes the set of nodes between the first and last excited lattice sites. We define
the width $W$ of the solution as the number of elements of the set $S$. 
Parameters $A$ and $B_{n}$ represent the amplitudes,
and $\eta $ is the decay rate. For solutions with odd $W$ (site-centered
configurations) the position
parameter is $n_{0}=0$, and for even $W$ (bond-centered
configurations) it is $n_{0}=0.5$. Independent
amplitudes $B_{n}$ are necessary to describe the core of the multi-humped
solutions accurately.

Since this ansatz is genuinely discrete and 
includes a purely exponential tail, the corresponding
approximations will only be valid for a small coupling strength,
since the solutions become increasingly 
smooth (sech-like) as the continuum limit
is approached. 
On the other hand, solutions with a
single excited site in the AC limit are only weakly influenced by the
extended coupling. For this reason, we aim to apply the VA to the
simplest solution type that ``feels" the extended coupling, which
corresponds to $W=3$. The substitution of this ansatz into the
Lagrangian and calculation of the ensuing sums yields the
effective Lagrangian
\begin{eqnarray}
\mathcal{L}_{\mathrm{eff}}&=&\frac{A^{4}}{2}E_{0,2\eta}
-A^{2}E_{0,\eta}+\sum_{m\in {\mathbb{N}}}\epsilon A^{2}E_{m,\eta}k_{|m|}\\[0.0ex]
\notag
&&
-\sum_{j\in S}
\left[ B_{j}^{2}(1-B_{j}^{2}/2)-\epsilon \sum_{n=-W}^{W}k_{j}\psi
_{n}^{\ast }\psi _{j}\right] ,  \label{eq:effLag}
\end{eqnarray}
where
\begin{equation}
E_{j,\eta}\equiv \frac{e^{-\eta (2+|j|)}}{e^{2\eta }-1}.
\end{equation}
The Euler-Lagrange equations derived from this Lagrangian are
\begin{equation}
\frac{\partial \mathcal{L}_{\mathrm{eff}}}{\partial p_{i}}=0,  \label{eq:EL}
\end{equation}
where $p_{i}\in\{A,B_n\}$ are parameters of the ansatz. The equation corresponding to
varying amplitude $A$ yields
\begin{eqnarray}
\notag
& 2A^{3}E_{0,2\eta}-2AE_{0,\eta}+\epsilon 2A\sum_{m\in {\mathbb{N}}}
E_{m,\eta}k_{m}+e^{-3\eta }\epsilon (B_{-1}+B_{1})k_{2} 
\\
\label{eq:A}
& +2e^{-2\eta }\epsilon (2B_{0}k_{2}+(B_{-1}+B_{1})k_{1})=0,
\end{eqnarray}
and the variation of the $B_{j}$'s yields,
\begin{eqnarray}
B_{-1}(B_{-1}^{2}-1)+\epsilon \left( Ae^{-3\eta }k_{2}+Ae^{-2\eta }k_{1}+{%
\sum_{m\in {\mathbb{N}}}}^{\prime }k_{|m|}B_{|m|-1}\right)  &=&0,
\label{eq:B2} \\
B_{0}(B_{0}^{2}-1)+\epsilon \left( 2Ae^{-2\eta
}k_{2}+k_{0}B_{0}+k_{1}(B_{-1}+B_{1})\right)  &=&0, \\
B_{1}(B_{1}^{2}-1)+\epsilon \left( Ae^{-3\eta }k_{2}+Ae^{-2\eta }k_{1}+{%
\sum_{m\in {\mathbb{N}}}}^{\prime }k_{|m|}B_{1-|m|}\right)  &=&0,
\label{eq:B3}
\end{eqnarray}
where the prime over the sum indicates that the $m=0$ entry is to be doubled.

Rather than performing the variation with respect to decay constant $\eta $,
we replace it by $\eta \equiv \ln \lambda $, where $\lambda $ is the
corresponding multiplier determined by the dynamical reduction.
Furthermore, we set $n_{0}=0$ rather than using it as a variational
parameter. This excludes asymmetric solutions, hence we can also set $%
B_{1}=B_{-1}$, further reducing the number of parameters.

\begin{figure}[tbh]
\centerline{
 \epsfig{file=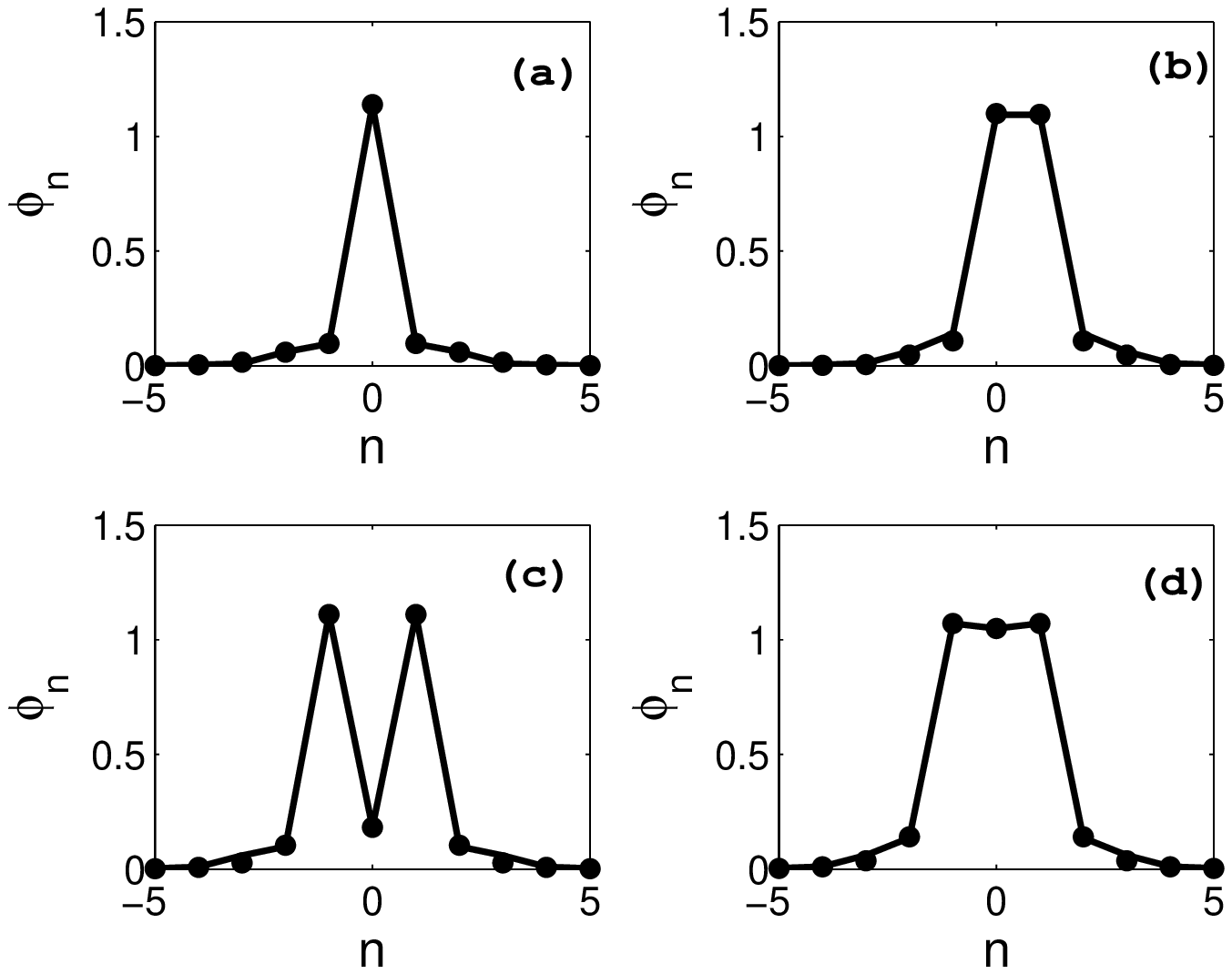,width=8 cm,angle=0}
}
\centerline{
 \epsfig{file=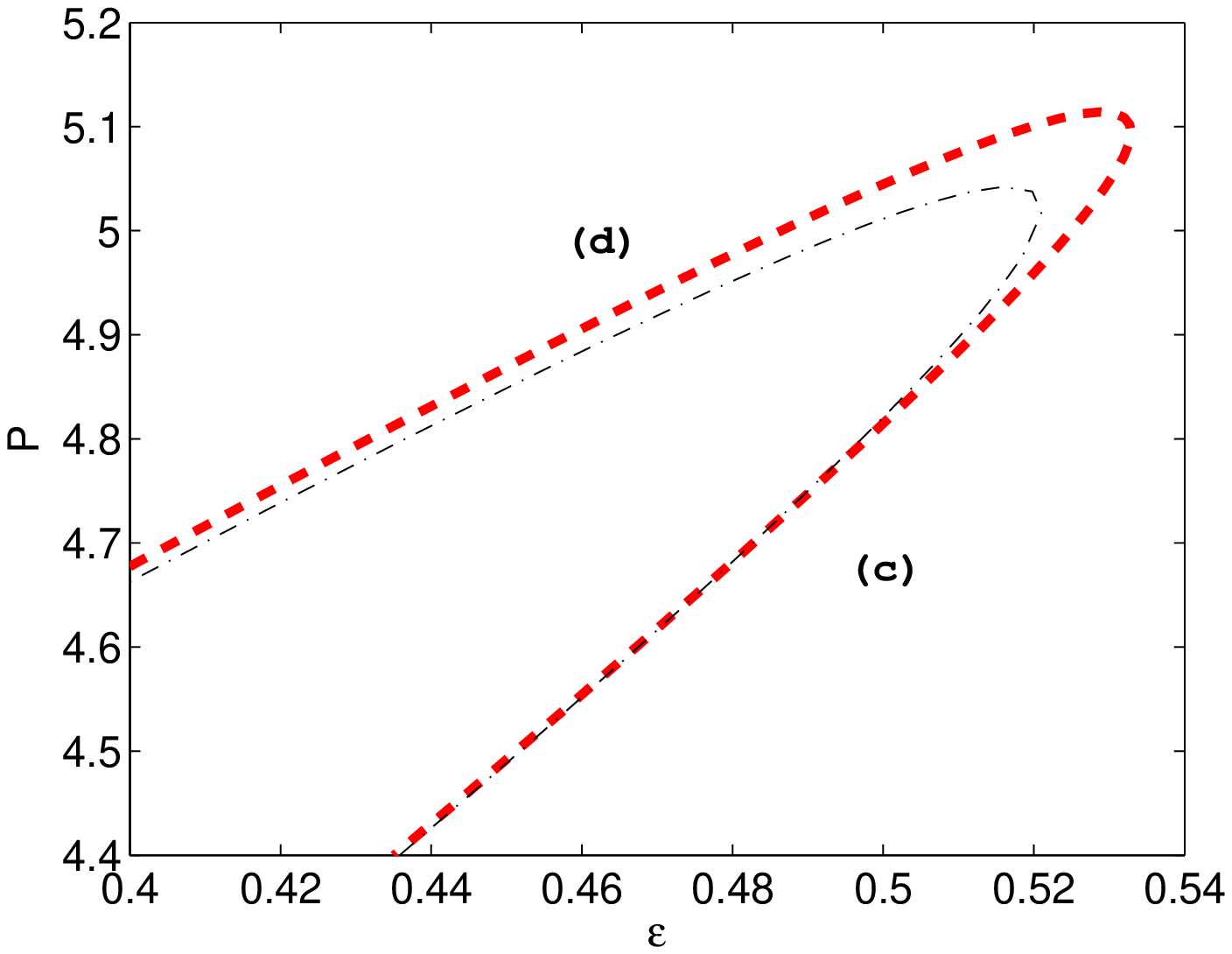,width=8 cm,angle=0}
 }
\centerline{
 \epsfig{file=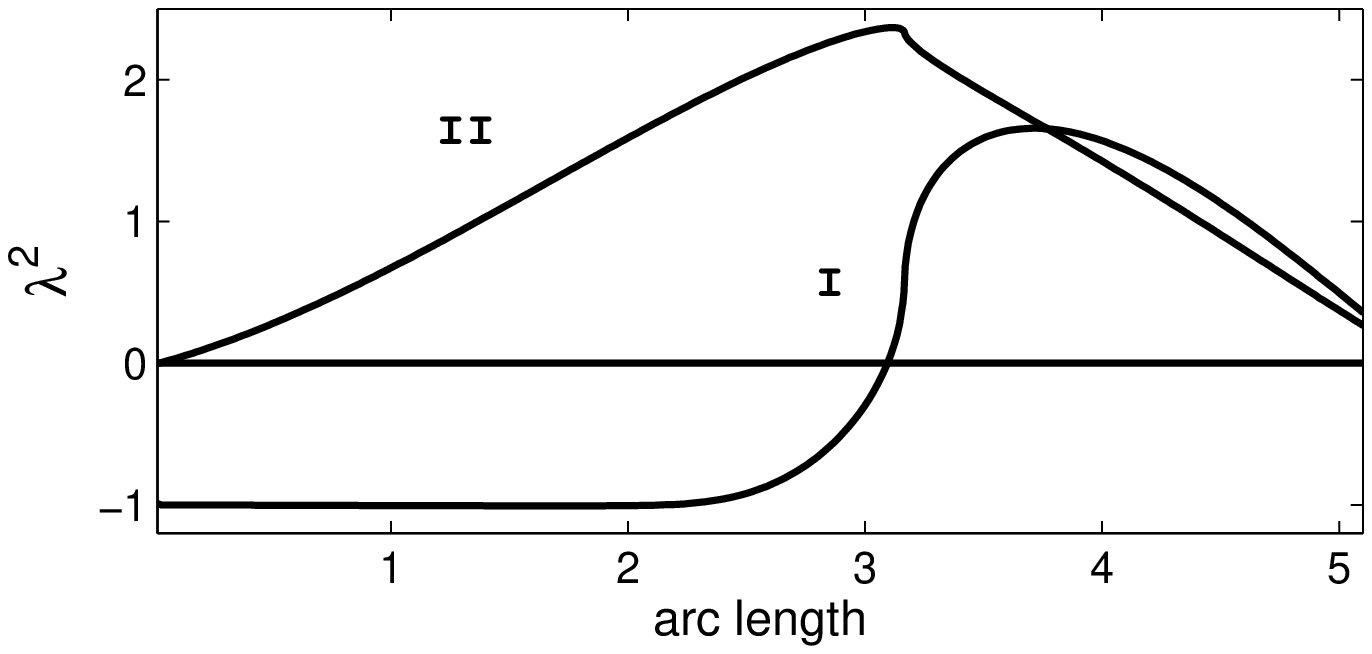,width=8 cm,angle=0}
 }
\caption{
Top panels: Numerical solutions of Eq.~\eqref{eq:dnls} with trivial
phase distributions (solid lines) and the variational solutions based on
ansatz \eqref{eq:ansatz} continued to $\protect\epsilon =0.1$ (markers).
Labels (a)--(d) correspond to those in Table~\protect\ref{tab:profiles}. The
error $\Vert \protect\phi -\protect\psi \Vert _{l^{2}}$ at these parameter
values are (a) $7.4\times 10^{-5}$ , (b) .0009, (c) .002, and (d) .001.
Middle panel: Zoom of the saddle-node bifurcation involving branches (c) and (d)
from the bottom panel of Fig.~\protect\ref{fig:bif}. The variational solution
(the black dash-dotted line) predicts the bifurcation at a slightly
smaller value of the coupling strength than the numerical solution (the red
dashed line). Both solutions are unstable in this parameter region. 
Bottom panel: Plot of two pairs of isolated eigenvalues versus the arc length
of the $P(\epsilon)$ curve of the middle panel. Here, zero arc length
corresponds to the (c) solution at $\epsilon=0$  
and the end of the arc length curve corresponds to the (d) solution at
$\epsilon=0$.
The eigenvalue pair (I) emerges from the edge of the 
continuous spectrum, i.e. $\lambda^2 = -1$. The eigenvalue pair (II)
is responsible for the instability
since it has positive real part (i.e. $\lambda^2 > 0$ ) in the entire region plotted.
}
\label{fig:compare_triv}
\end{figure}

Equations (\ref{eq:B2})--(\ref{eq:B3}) pertain to the localized pattern with
$W=3$, but they can be readily extended to other cases. Using these
equations, we can approximate all the solutions with trivial phases from
Fig.~\ref{fig:bif}, i.e., branches (a)--(e), including the single- and
double-site solutions, see the top panel of Fig.~\ref{fig:compare_triv}.
Indeed, the predictions are so good that one cannot see the differences when
the bifurcation diagram based on the VA is juxtaposed with Fig.~\ref{fig:bif}.
A zoom of the saddle-node bifurcation between solutions (c) and (d) is
shown in the bottom panel of Fig.~\ref{fig:compare_triv} where the
differences are visible.
One might expect the (c) and (d) solutions to exchange stability, 
which would be the case in a standard saddle-node
bifurcation in a low-dimensional model. However, due to the higher
dimensionality of the DNLS model considered here, there will be additional eigenvalues
that could affect the stability character of the solution. Indeed,
for the (c) and (d) solutions there are two pairs of isolated eigenvalues
(and one pair at the origin).
One of these pairs moves along the imaginary axis and becomes real
(see curve (I) of the bottom panel of Fig.~\ref{fig:compare_triv})
for some critical value of $\epsilon$. However, due to the existence
of the other pair of eigenvalues (see curve (II)), which
has non-zero real part in the parameter region shown, 
both solutions are unstable.

It is obvious that ansatz \eqref{eq:ansatz} can only capture solutions
with the trivial phase structure, as it is real-valued. Nonetheless, it is
informative to inspect the AC limit in the framework of the variational
equations to see what types of solutions are candidates to be approximated.
With $\epsilon =0$ the equations reduce to $A=0$ and $B_{j}(B_{j}^{2}-1)=0$
such that the corresponding VA solutions coincide exactly with solutions of
the full problem \eqref{eq:dnls-sta} with the phases given by either $0$
or $\pm \pi $, as expected. The fact that the VA and full solutions match at
$\epsilon =0$ follows from the choice of the ansatz. This is not the case if
the standard ansatz, based on the exponential cusp with the single central
point is used, cf. Ref.~\cite{MM91}.

\subsection{Nontrivial Phase Distributions}

\begin{figure}[thbp]
\centerline{
 \epsfig{file=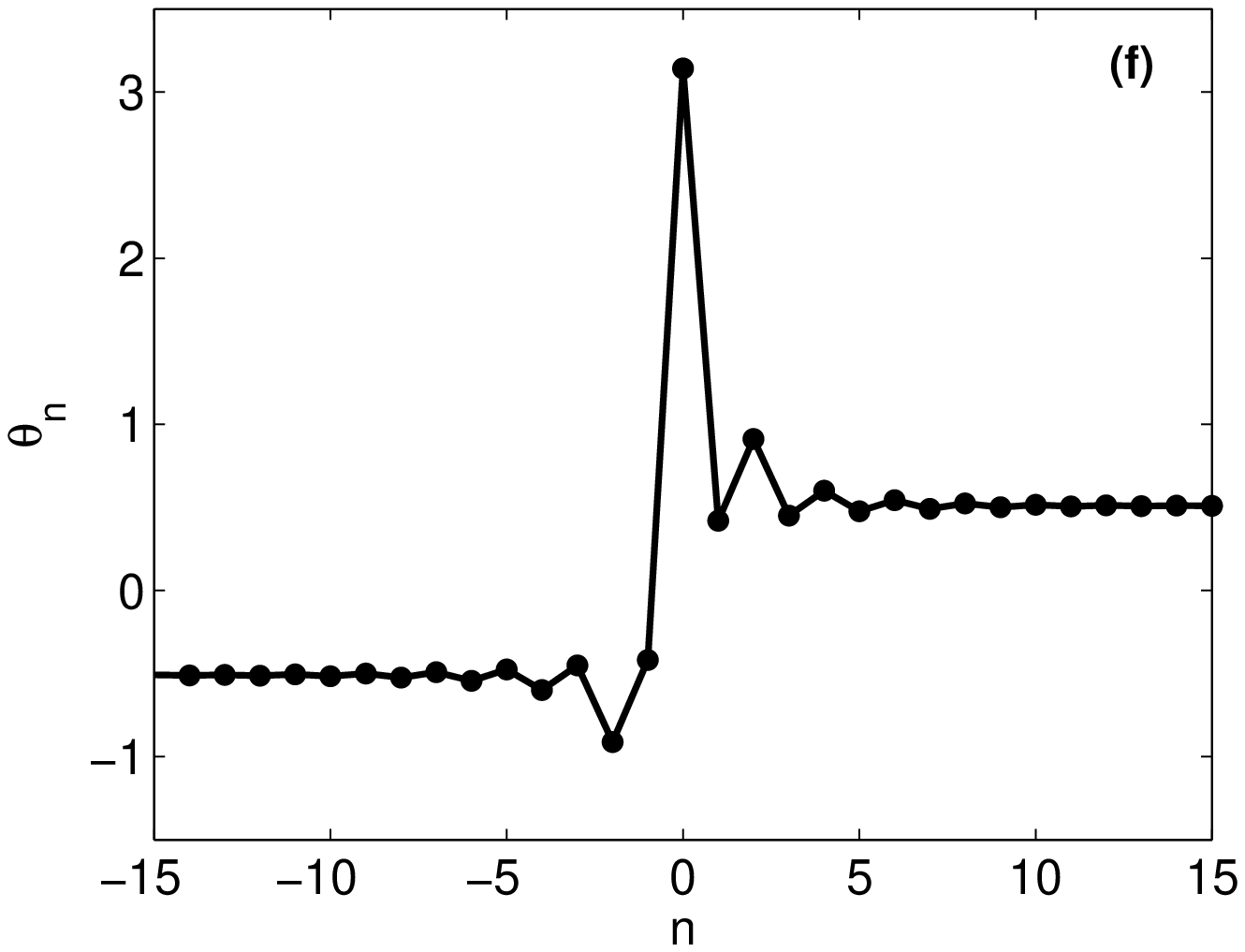,width=6.0 cm,angle=0}
 \epsfig{file=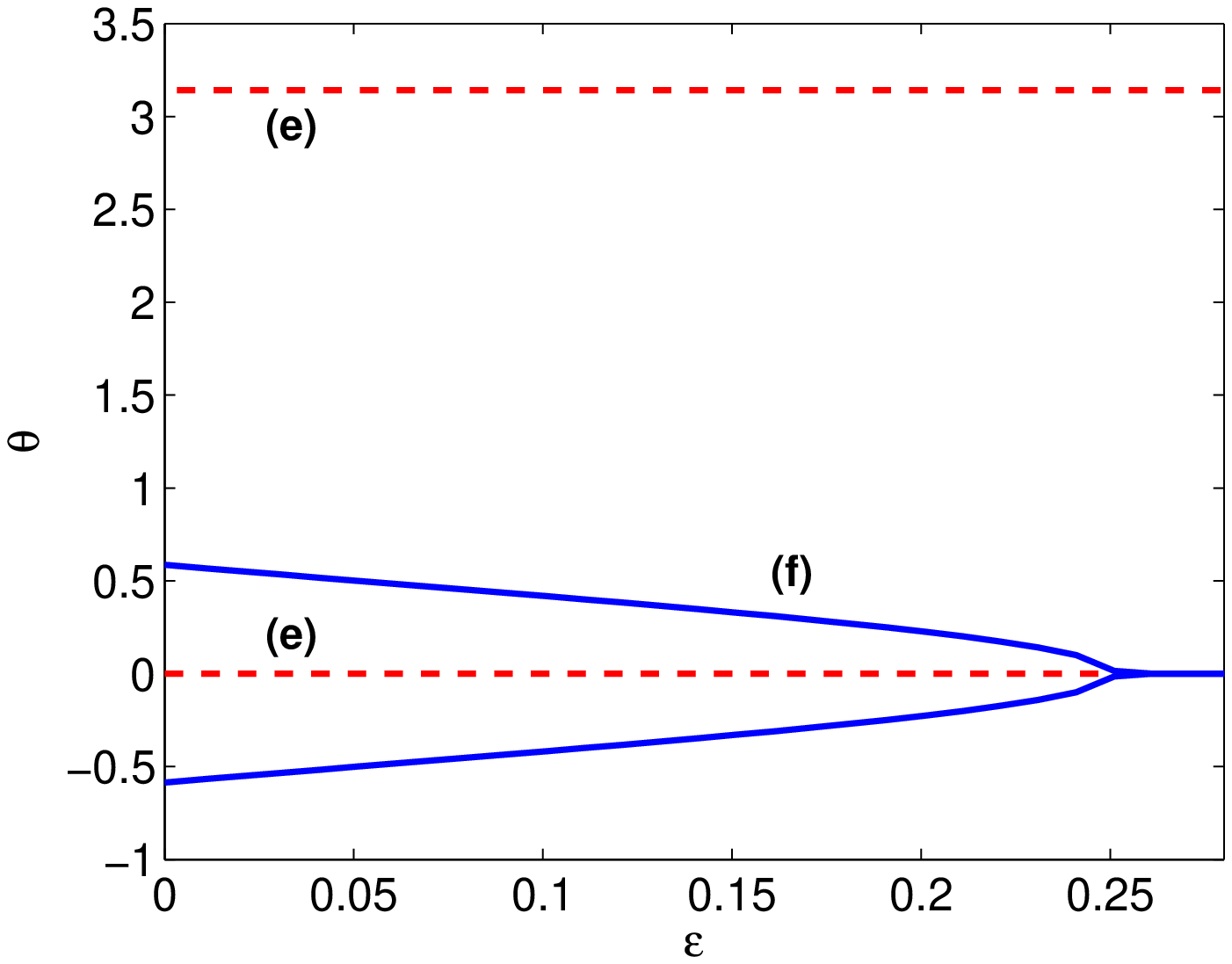,width=6.0 cm,angle=0}
 \epsfig{file=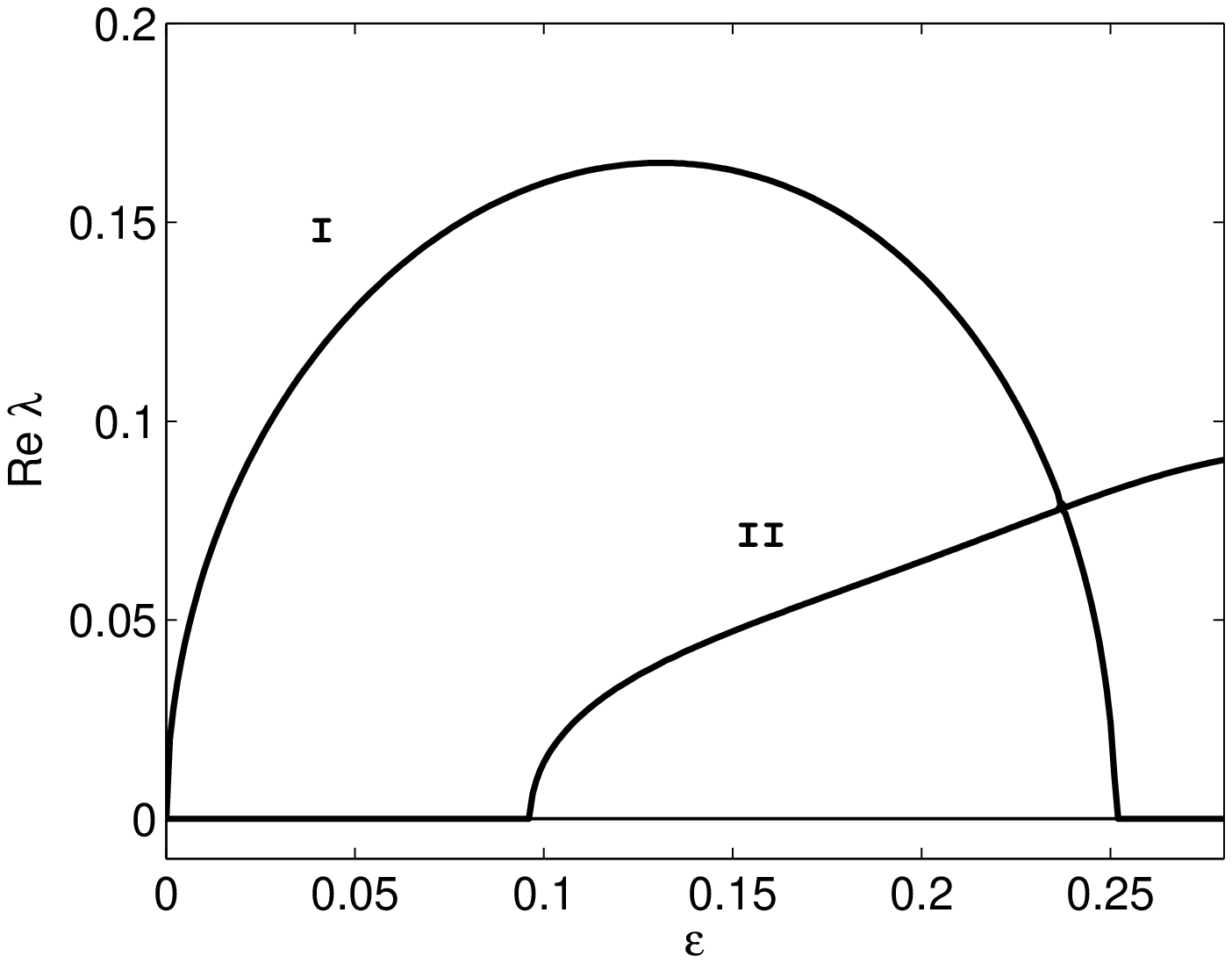,width=6.0 cm,angle=0}
 }
\caption{(Color online)
Left:
Phases of the nontrivial solution (f) for $\epsilon=0.1$ and the same parameter
values as in Table~\protect\ref{tab:profiles}. 
Middle: Phases at the $n=\pm 1$ and $n=0$ sites for versus $\epsilon$. 
Solution (e) corresponds to $\theta_{-1}=\theta_{1}=0$ and $\theta_0 = \pi$
(red dashed lines)
while solution (f) corresponds $\theta_0=\pi$ and
nontrivial $\theta_{\pm 1}$ 
(blue solid curves).
%
The solutions (e) and (f)
collide at the critical point, $\protect\epsilon \approx 0.25$,
where the latter terminates through the ensuing subcritical pitchfork
bifurcation. 
Right:
The two eigenvalues with the largest real part, found from a solution of 
Eq.~\eqref{eq:eig} corresponding to solution (e). The eigenvalue corresponding
to the bifurcation shown in the middle panel is labeled I. However, due to
the emergence of a second real eigenvalue pair (II), the 
solution (e) is unstable
in all of the parameter region shown. The corresponding spectral picture for
the solution (f) is similar, but without the 
curve I. Thus, solution (f) becomes
unstable at $\protect\epsilon \approx 0.098$. }
\label{fig:phases}
\end{figure}

In order to capture solutions with nontrivial phases, phase must be added to
the ansatz. To motivate our choice, we look closer at the solutions with nontrivial
phases. Similarly to the trivial-phase ansatz \eqref{eq:ansatz}, we separate the
initially excited sites from the others. The outer part of the solution
should have exponential decay, but with a varying phase. In the
left
panel of Fig.~\ref{fig:phases} the phases of the numerically found
nontrivial solution (f) for $\epsilon=0.1$ 
are plotted against the lattice coordinate. A precise
description of each phase would make the ansatz too complex, leading to
intractable sums in the effective Lagrangian. However, it is clear that the
solution will be an exponentially localized one, and a coarse approximation
for the phases may be sufficient. 
Therefore, motivated by Fig.~\ref{fig:phases}, 
we assume a constant phase, $\kappa _{-1}$, for $n<0$ and
another constant phase, $\kappa _{1}$, for $n>0$. 
For the core of the
solution, one might introduce the number of additional phase parameters
equal to $W$, one per each site, but, aiming to keep the number of
parameters reasonably low, we make the following observations
based on the dependence of the phases for the core sites as
a function of $\epsilon$ (see middle panel of Fig.~\ref{fig:phases}).
We have observed that the phase at
the central site ($n=0$) stays almost constant as the coupling is turned on
and the difference between the other points in the core ($n=\pm 1$) are
$\theta _{\pm 1}(\epsilon )\approx \theta _{\pm 1}^{0}\pm b$, where $\theta
_{n}^{0}$ is the phase at the $n$-th site in the AC limit, and $b$ is a
constant which depends on parameters of the system (see the middle panel of
Fig.~\ref{fig:phases}). Therefore, for the nontrivial-phase solution with
width $W=3$ we adopt the following ansatz,
\begin{equation}
\psi _{n}=\left\{
\begin{array}{ll}
A\exp (\mathrm{i}\kappa _{-1})\exp (-\eta |n|)&\mathrm{~for~}  n<-1, \\[1.0ex]
B\exp (\mathrm{i}(\phi _{-1}^{0}-b))&\mathrm{~for~} n=-1, \\[1.0ex]
C\exp (\mathrm{i}\phi _{0}^{0})&\mathrm{~for~} n=0, \\[1.0ex]
B\exp (\mathrm{i}(\phi _{1}^{0}+b))&\mathrm{~for~} n=1, \\[1.0ex]
A\exp (\mathrm{i}\kappa _{1})\exp (-\eta |n|)&\mathrm{~for~} n>1,%
\end{array}%
\right.\label{eq:ansatz-nontriv}
\end{equation}
where real parameters $A,B,C$ represent the amplitude, and the phases are
represented by $\kappa _{\pm 1}$ and $b$. As before, $\eta $ is determined
via the dynamical reduction.
We also present in the right panel of Fig.~\ref{fig:phases} the instability
eigenvalues for the trivial phase solution (e). As can be seen
from this panel, if it was not for the eigenvalue pair (II), this solution
would gain stability after collision with the nontrivial phase
solution (f) for $\epsilon>0.25$. 
In particular, this eigenvalue
plot in conjunction with the middle panel of the figure illustrating
the phases of the different branches clearly showcases the existence
of a subcritical pitchfork bifurcation, which leads to the termination
of the nontrivial phase branch (f).

The effective Lagrangian corresponding to the nontrivial-phase ansatz is the
same as in the case of the solution with the trivial phases, with the
exception that $\psi $ appearing in Eq.~\eqref{eq:EL} will be the one
of Eq.~\eqref{eq:ansatz-nontriv} and with $B_{\pm 1}=B$ and $B_{0}=C$.

\begin{figure}[t]
\centerline{
 \epsfig{file=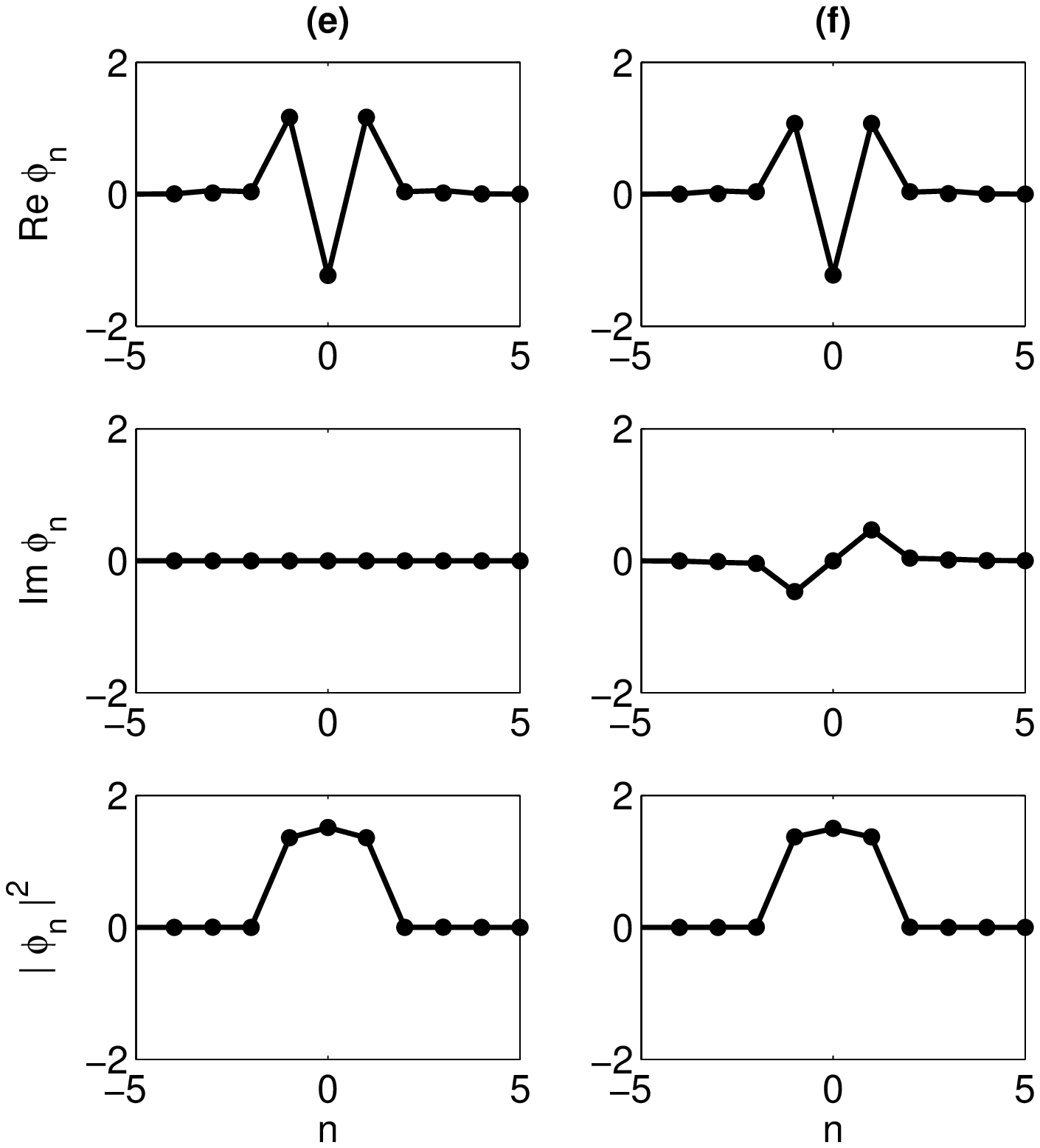,width=8 cm,angle=0}
}
\centerline{
 \epsfig{file=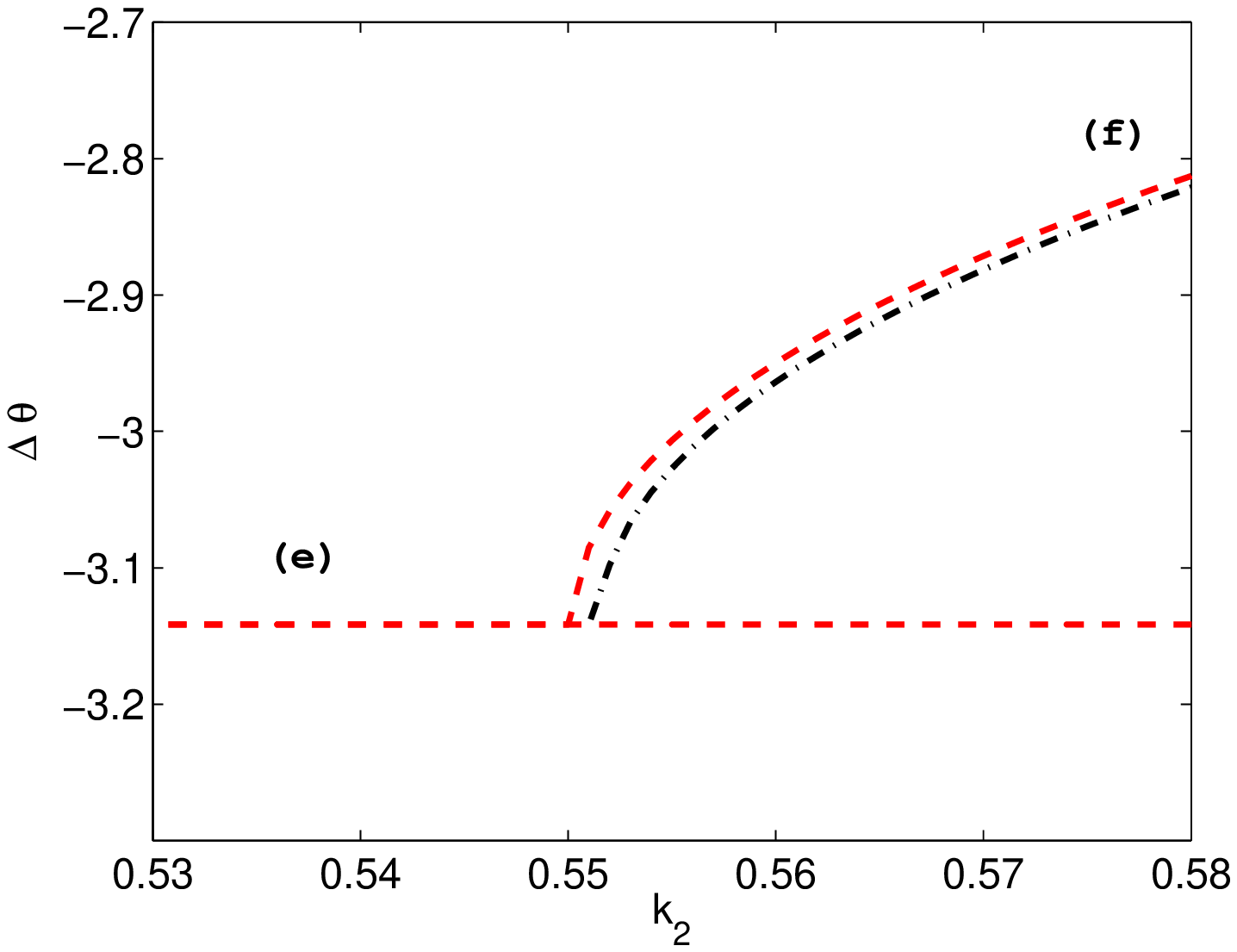,width=8 cm,angle=0}
}
\caption{
Top panels: The real part (top row) and imaginary part (middle row) of
the trivial-phase solution (e) and the one with the nontrivial phase (f).
The numerically exact solutions (solid lines) are close to their variational
counterparts (markers) based on ansatz \eqref{eq:ansatz-nontriv}, with
errors $\Vert \protect\phi -\protect\psi \Vert _{l^{2}}$ being .004 (e) and
.004 (f). The parameters are again the same as in 
Table~\ref{tab:profiles} and $\epsilon=1$. 
Bottom panel: Phase difference $\Delta \protect\theta =\protect%
\theta _{0}-\protect\theta _{1}$ for the numerical (red dashed curve)
and variational (black dash-dotted curve) solutions
for branches (e) and (f) versus the
outer-coupling parameter $k_{2}$ for $\epsilon =0.1$. The
bifurcation here is of the pitchfork type, with the mirror-symmetric partner
of (f) connecting with solution (e) as well.}
\label{fig:compare_nontriv}
\end{figure}

The variation of the effective Lagrangian with respect to the
parameters $A,B,C,\kappa _{-1},\kappa _{1}$ and $b$ yields, respectively,
%
\begin{eqnarray}
\notag
0 &=&-2AE_{0,\eta}+B\epsilon (k_{1}e^{-2\eta }+k_{2}e^{-3\eta })(\cos
(r^{-})+\cos (r^{+}))+\epsilon e^{-2\eta }Ck_{2}(\cos (\kappa _{-1}-\theta
_{0}^{0})+\cos (\kappa _{1}-\theta _{0}^{0}))\\
\label{eq:A-nontriv}
&&+\epsilon 2A\sum_{m\in S}E_{m,\eta}k_{m}+2A^{3}E_{0,2\eta}, \\[1.0ex]
\notag
0 &=&\epsilon Bk_{2}\cos (\theta _{-1}^{0}+2b-\theta _{1}^{0})+\epsilon
(Ck_{1}(\cos (s^{-})+\cos (s^{+}))+A(\cos (r^{-})+\cos
(r^{+}))(k_{1}e^{-2\eta }+k_{2}e^{-3\eta }))  \\
\label{eq:B-nontriv} 
&&
+B(B^{2}-1)+\epsilon Bk_{0},  \\[1.0ex]
\label{eq:C-nontriv} 
0 &=&\epsilon (Ck_{0}+Bk_{1}(\cos (s^{-})+\cos (s^{+}))+Ak_{2}e^{-2\eta
}(\cos (\theta _{0}^{0}+\kappa _{-1})+\cos (-\theta _{0}^{0}+\kappa _{1})))
+C(C^{2}-1), \\[1.0ex]
0 &=&\epsilon A(Ck_{2}e^{-2\eta }\sin (\kappa _{-1}-\theta _{0}^{0})+B\sin
(r^{-})(k_{1}e^{-2\eta }+k_{2}e^{-3\eta })),  \label{eq:kl} \\[1.0ex]
0 &=&\epsilon A(Ck_{2}e^{-2\eta }\sin (\kappa _{1}-\theta _{0}^{0})+B\sin
(r^{+})(k_{1}e^{-2\eta }+k_{2}e^{-3\eta })),  \label{eq:kr} \\[1.0ex]
0 &=&\epsilon B(2A(\sin (r^{+})-\sin (r^{-}))(k_{1}e^{-2\eta
}+k_{2}e^{-3\eta })-4k_{2}B\sin (\theta _{0}^{0}+2b-\theta _{-1}^{0})) \notag  \\
&&-\epsilon BCk_{1}(\sin (s^{+})-\sin (s^{-})), \label{eq:b}
\end{eqnarray}
%
where $r^{-}\equiv \kappa _{-1}-b-\theta _{-1}^{0}$,$~r^{+}\equiv \kappa
_{1}+b-\theta _{1}^{0}$, $s^{-}=-\theta _{-1}^{0}-b+\theta _{0}^{0}$, and $%
s^{+}=-\theta _{1}^{0}+b+\theta _{0}^{0}$. These equations reduce to those
for the trivial-phase solutions, displayed in the previous section for $%
\kappa _{\pm 1}=0$ and $b=\theta _{1}^{0}$. To look at the bifurcation
between the trivial- and nontrivial-phase solutions (e) and (f) in another
way, we can vary the outer coupling parameter $k_{2}$ in 
Eq.~(\ref{eq:dnls-nnn}), 
while keeping $\epsilon $ fixed. As mentioned before, it is
easier to identify the bifurcation through the comparison of phases,
therefore we consider the phase difference $\Delta \theta =\theta
_{0}-\theta _{1}$, which has the advantage of being independent of any
constant phase (since the solutions are gauge invariant).
The agreement between the
numerically exact solution and the one based on the VA is quite remarkable,
see the bottom panel of Fig.~\ref{fig:compare_nontriv}. In this panel,
the mirror symmetric nontrivial phase solution (i.e., the one with
opposite relative phases) has been omitted.

%
Using these variational equations, it is possible to approximate the
bifurcation point in the bottom panel of 
Fig.~\ref{fig:compare_nontriv} which connects solutions
(e) and (f) without actually solving Eqs.~\eqref{eq:A-nontriv}--\eqref{eq:b}. 
Making use of the approximations $\kappa _{1}=-\kappa _{-1}$ and $%
\kappa _{-1}=(b+\theta _{-1})/2$ and Taylor-expanding Eq.~\eqref{eq:kl}
leads to
\begin{equation}
\kappa _{-1}\approx \pm \sqrt{\frac{B(k_{1}e^{-2\eta }+k_{2}e^{-3\eta
})-2Ck_{2}e^{-2\eta }}{B(k_{1}e^{-2\eta }+k_{2}e^{-3\eta
})/12-2Ck_{2}e^{-2\eta }/96}},  \label{eq:akl}
\end{equation}
or $\kappa _{-1}=0$. Using values of $B$ and $C$ obtained from the (much
simpler) equations \eqref{eq:A}--\eqref{eq:B3}, we thus produce an
accurate prediction of the bifurcation, avoiding the need to solve the full
system~of Eqs.~\eqref{eq:A-nontriv}--\eqref{eq:b}, since all terms in 
Eq.~\eqref{eq:akl} are known. The right hand side
of equation~\eqref{eq:akl} vanishes at $%
k_{2}\approx 0.538$, which is close the actual bifurcation value of $%
k_{2}\approx 0.551$ in Fig.~\ref{fig:compare_nontriv}. This is an
improvement in comparison to the prediction based on Eq.~\eqref{eq:cond},
which is $k_{2}\approx 0.5$. The deviation from the latter simple
leading order
prediction can be justified by the use of $\epsilon=1$, whereas the
derivation of \cite{PKnnn} was valid for small 
$\epsilon$.

%

\begin{figure}[t]
\centerline{
 \epsfig{file=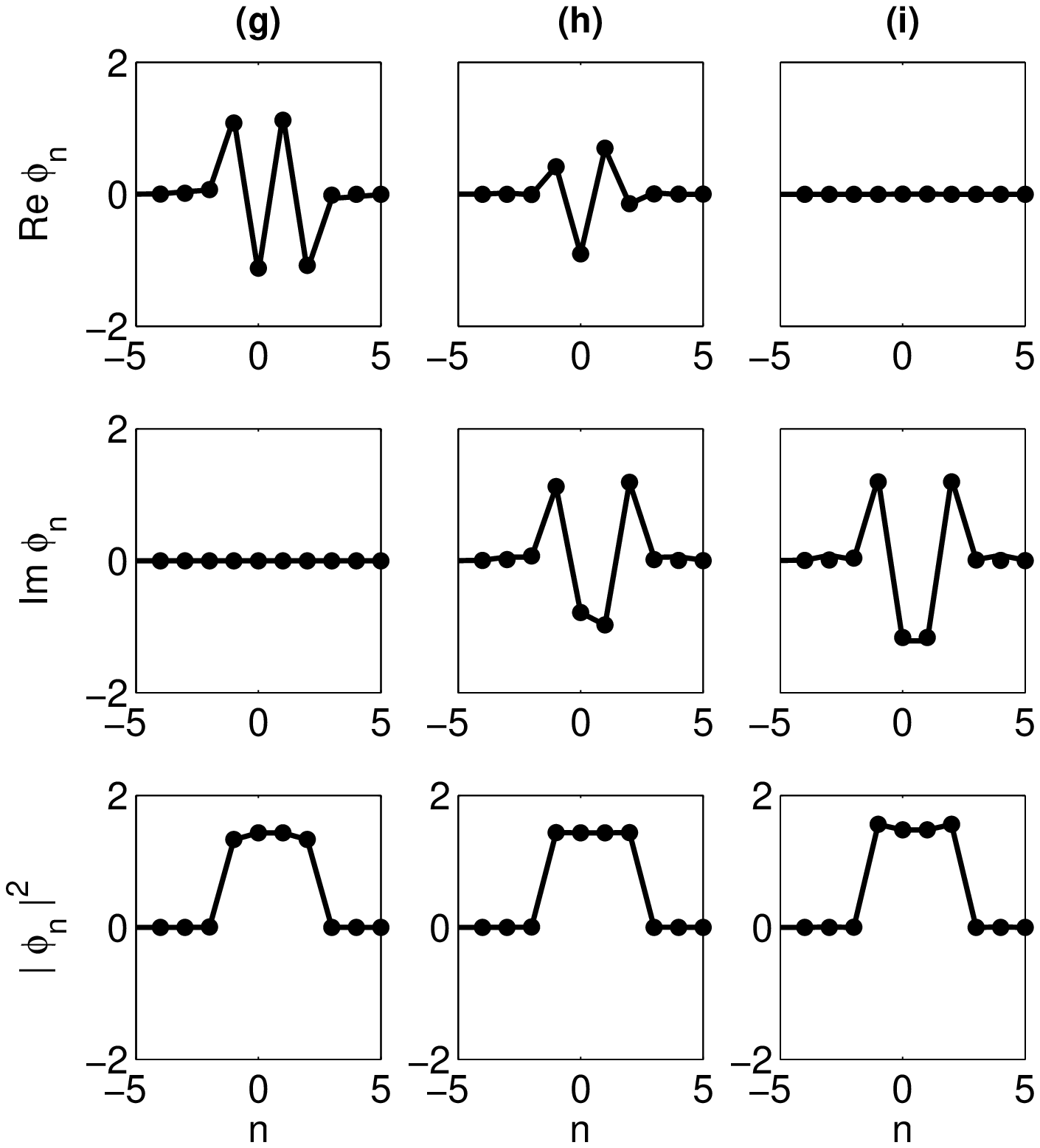,width=8 cm,angle=0}
}
\centerline{
 \epsfig{file=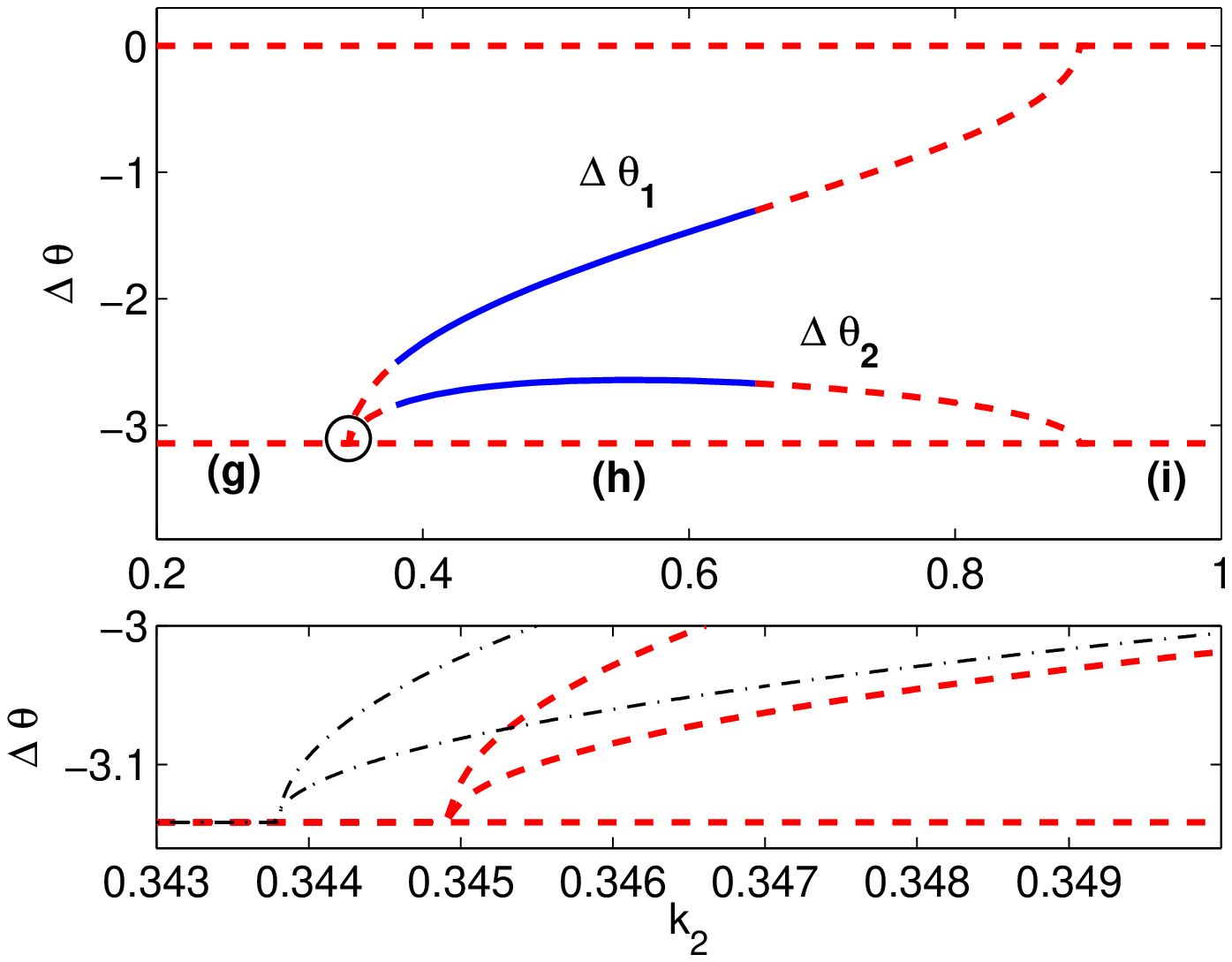,width=8 cm,angle=0}
}
\caption{(Color online)
Top six panels: Profiles of the four-site trivial-phase solutions (g) and
(i), which are connected through the nontrivial phase solution (h). 
Bottom two panels: Plots
of the phases differences $\Delta \protect\theta _{1}=\protect\theta _{1}-%
\protect\theta _{0}$ and $\Delta \protect\theta _{2}=\protect\theta _{2}-%
\protect\theta _{1}$ of solutions (g)--(i). At the resolution
of the top subpanel the
difference between the variational and numerical curves cannot be seen.
Solution (h) undergoes two stability changes in the parameter region shown
here (dashed red and blue solid lines show the unstable and stable
families, respectively). 
Bottom subpanel: Zoom of the circled area 
showing the discrepancy between the variational approximation (black
dash-dotted lines) and numerically generated curves (red dashed lines). }
\label{fig:compare_nontriv_four}
\end{figure}

As a final example, we will briefly consider a four-site solution $(W=4)$.
The corresponding even counterpart of ansatz~\eqref{eq:ansatz-nontriv} is
\begin{equation}
\psi _{n}=\left\{
\begin{array}{ll}
A\exp (\mathrm{i}\kappa _{-1})\exp (-\eta |n+n_{0}|) & {\rm~for~} n<-1 \\[1.0ex]
B\exp (\mathrm{i}(\phi _{-1}^{0}-b)) & {\rm~for~} n=-1 \\[1.0ex]
C\exp (\mathrm{i}(\phi _{0}^{0}-c)) & {\rm~for~} n=0 \\[1.0ex]
C\exp (\mathrm{i}(\phi _{1}^{0}+c)) & {\rm~for~} n=1 \\[1.0ex]
B\exp (\mathrm{i}(\phi _{1}^{0}+b)) & {\rm~for~} n=2 \\[1.0ex]
A\exp (\mathrm{i}\kappa _{1})\exp (-\eta |n+n_{0}|) & {\rm~for~} n>2,%
\end{array}%
\right. \label{eq:ansatz-nontriv4}
\end{equation}%
where $n_{0}=0.5$. The effective Lagrangian corresponding to ansatz
\eqref{eq:ansatz-nontriv4} is the same as the trivial-phase one, with the
exception that once again the 
$\psi $ appearing in Eq.~\eqref{eq:EL} will be of the form of
Eq.~\eqref{eq:ansatz-nontriv4}, with $B_{-1}=B_{2}=B$, $B_{0}=B_{1}=C$, and
\begin{equation}
E_{j,\eta}=\frac{e^{-\eta (2+|j|+2n_{0})}+e^{-\eta (4+|j|+2n_{0})}}{e^{2\eta }-1}.
\end{equation}
The extra term $e^{-\eta (4+|j|+2n_{0})}$ appearing in the expression for $%
E_{j,\eta}$ for the solutions with the even width is due to the fact the
summation in the Lagrangian is no longer symmetric about the zero site. The
variations with respect to the parameters $A,B,C,\kappa _{-1},\kappa _{1},b$
and $c$ yield, respectively,
%
%
\begin{eqnarray}
0 &=&A^{3}E_{0,2\eta}-AE_{0,\eta}+\epsilon A\sum_{m\in
S}E_{m,\eta}k_{m}+\epsilon B\left( k_{1}e^{-\eta (2+n_{0})}+k_{2}e^{-\eta
(3+n_{0})}\right) \cos (r_{b}^{-})  \label{eq:A-nontriv4} \\
&&+\epsilon B\left( k_{1}e^{-\eta (3+n_{0})}+k_{2}e^{-\eta (4+n_{0})}\right)
\cos (r_{b}^{+})+\epsilon Ck_{2}e^{-\eta (2+n_{0})}\cos (r_{c}^{-})+\epsilon
Ck_{2}e^{-\eta (3+n_{0})}\cos (r_{c}^{+}),  \notag \\[2ex]
0 &=&2B(B^{2}-1)+2\epsilon Bk_{0}+\epsilon Ck_{1}(\cos
(s_{c}^{-})+\cos (s_{c}^{+}))+\epsilon Ck_{2}(\cos (s_{b}^{-})+\cos
(s_{b}^{+}))  \label{eq:B-nontriv4} \\
&&+\epsilon A\left( \cos (r_{b}^{-})k_{1}e^{-\eta (2+n_{0})}+\cos
(r_{b}^{-})k_{2}e^{-\eta (3+n_{0})}+\cos (r_{b}^{+})k_{1}e^{-\eta
(3+n_{0})}+\cos (r_{b}^{+})k_{2}e^{-\eta (4+n_{0})}\right) ,  \notag \\[2ex]
0 &=&2C(C^{2}-1)+\epsilon B(k_{1}\cos (s_{c}^{-})+k_{1}\cos
(s_{c}^{+})+k_{2}\cos (s_{b}^{-})+k_{2}\cos (s_{b}^{+}))+2\epsilon Ck_{0}
\label{eq:C-nontriv4} \\
&&+2\epsilon Ck_{1}\cos (\theta _{0}+2c-\theta _{1})+\epsilon
Ak_{2}(e^{-\eta (2+n_{0})}\cos (\theta _{0}-\kappa _{-1}+c)+e^{-\eta
(3+n_{0})}\cos (\theta _{1}-\kappa _{1}-c)),  \notag \\[2ex]
0 &=&\epsilon A(Ck_{2}e^{-\eta (2+n_{0}))}\sin (\kappa _{-1}-\theta
_{0}^{0}-c)+B\sin (r_{b}^{-})(k_{1}e^{-\eta (2+n_{0})}+k_{2}e^{-\eta
(3+n_{0})}),  \label{eq:kl4} \\[2ex]
0 &=&\epsilon A(Ck_{2}e^{-\eta (3+n_{0})}\sin (\kappa _{1}-\theta
_{1}+c)+B\sin (r_{b}^{+})(k_{1}e^{-\eta (3+n_{0})}+k_{2}e^{-\eta (4+n_{0})}),
\label{eq:kr4} \\[2ex]
0 &=&-\epsilon BC(k_{2}\sin (s_{b}^{-})+k_{2}\sin (s_{b}^{+})+k_{1}\sin
(s_{c}^{+})+k_{1}\sin (s_{c}^{-}))  \label{eq:b4} \\
&&+\epsilon BA\left( \sin (r_{b}^{-})k_{2}e^{-\eta (3+n_{0})}-\sin
(r_{b}^{+})k_{2}e^{-\eta (4+n_{0})}-\sin (r_{b}^{+})k_{1}e^{-\eta
(3+n_{0})}+\sin (r_{b}^{-})k_{1}e^{-\eta (2+n_{0})}\right) ,  \notag \\[2ex]
0 &=&-\epsilon CB(k_{2}\sin (s_{b}^{-})+k_{2}\sin (s_{b}^{+})-k_{1}\sin
(s_{c}^{+})-k_{1}\sin (s_{c}^{-}))-2k_{1}\epsilon C^{2}\sin (\theta
_{0}-2c-\theta _{1})  \label{eq:c4} \\
&&+\epsilon AC\left( \sin (r_{c}^{-})k_{2}e^{-\eta (2+n_{0})}-\sin
(r_{c}^{+})k_{2}e^{-\eta (3+n_{0})}\right) ,  \notag
\end{eqnarray}
%
where $r_{b}^{-}\equiv \kappa _{-1}-b-\theta _{-1}$, $r_{b}^{+}\equiv \kappa
_{1}+b-\theta _{2}$, $r_{c}^{-}\equiv \kappa _{-1}-c-\theta _{0}$ and $%
r_{c}^{+}\equiv \kappa _{1}+c-\theta _{1}$, $s_{b}^{-}\equiv \theta
_{-1}-\theta _{1}+c+b$, $s_{b}^{+}\equiv \theta _{0}-\theta _{2}+c+b$, $%
s_{c}^{-}\equiv \theta _{1}-\theta _{2}-c+b$, $s_{c}^{+}\equiv \theta
_{-1}-\theta _{0}-c+b$. Although it is not the goal of this work to provide
for a list of every possible bifurcation scenario, it is worth
mentioning that nontrivial-phase solutions connect various types of
phase-trivial ones. For example, as depicted in 
Fig.~\ref{fig:compare_nontriv_four}, the nontrivial solution (h) connects the
phase-trivial ones (g) and (i), with $k_{2}$ treated as the bifurcation
parameter. Solution (g), with phases differences $\{\Delta \theta
_{1},\Delta \theta _{2}\}=\{-\pi ,-\pi \}$, bifurcates at $k_{2}\approx 0.345
$, where the nontrivial-phase solution (h) emerges. The phase differences of
solution (h) change as $k_{2}$ varies. At $k_{2}\approx 0.894$, solution (h)
collides with and is annihilated by the trivial-phase solution (i), which
has phases differences $\{\Delta \theta _{1},\Delta \theta _{2}\}=\{0,-\pi \}
$. I.e., the phenomenology involves a supercritical pitchfork (in
the bifurcation parameter $k_2$) for the emergence of the nontrivial
phase branch (h) from (g) and a subcritical pitchfork which results
from the collision of (g) with (i).
This is similar to how asymmetric solutions connect solutions of varying
width in DNLS equations with higher-order nonlinearities 
\cite{CP09,Ricardo,OJE,TD} with the exception that the overall
(in)stability of the
trivial-phase solutions is not affected by collision with
nontrivial-phase 
solutions\footnote{Nevertheless, along the particular eigendirection of the 
bifurcation, stability is exchanged between these solutions, as is mandated by
the pitchfork character of the bifurcation.}. The nontrivial-phase
solution itself undergoes stability change, as can be seen in the
bottom panel of Fig.~\ref{fig:compare_nontriv_four}. 
The VA
captures this scenario remarkably well. Indeed, in the top subpanel
of the bottom panels in
Fig.~\ref{fig:compare_nontriv_four}, the difference between the numerical
and variational solutions cannot be spotted. Actually, a very strong zoom
around the bifurcation point is needed to depict the difference
(see bottom subpanel) and the relevant error in the identification
of the critical point is less than $0.4$ $\%$.

\section{Conclusions}\label{SEC:conc}

\label{sec:theend}

We have revisited discrete nonlinear Schr\"odinger (DNLS) 
models with extended linear couplings in the
lattice, developing a new version of the variational approximation (VA) for
the models of this type. Using an ansatz that coincides with exact solutions
in the anti-continuum limit, we were able to accurately describe, for
the first time, multi-humped solutions, including those with nontrivial
phase structures, and the bifurcations linking different species of the
discrete solitons. 
Pitchfork bifurcations connecting the solutions with the
nontrivial and trivial phase structures 
were identified, which the VA was successful in
predicting, thus demonstrating its reliability. In particular, the strength
of VA analysis is that it
provides simple formulas to predict such bifurcation points.
%
%
Like in other settings where the VA is used, a precise evaluation of the
validity of this approximation is an open question, therefore we relied on
the direct comparison with numerical solutions of the full problem. It was
found that the accuracy of the VA is quite good for small lattice coupling
strength.

There remain several other open problems concerning 
next-nearest-neighbor DNLS equations, such
as how to compute the stable manifolds in terms of the dynamical reduction,
or what the appropriate continuum-limit counterparts of these models are
and what modifications to the existence and spectral properties of the
solitary waves the long-range kernels may induce more generally.
As concerns the VA, its time-dependent
version can be also used to predict the linear stability of the solutions,
see Refs.~\cite{Ma02} and \cite{CP09}, which may be another direction for
the development of the analysis reported in this work. Finally, bearing
in mind some of the important consequences of long-range interactions
in higher dimensions, such as the stabilization of unstable
vortices among others \cite{wieslaw}, consideration of such questions
in the discrete realm is a theme of particular interest in its own
right.

\section*{Acknowledgments}

The authors would like to thank Guido Schneider and Faustino Palmero Acebedo for
useful discussions. C.C. and P.G.K. thank Vassilis Koukouloyannis
for numerous discussions and for sharing insights on corresponding phenomena in 
Klein-Gordon chains, especially for the suggestion to inspect the bifurcations
in $(k_2,\Delta \theta)$ space. The work of C.C. is partially supported by the Deutsche
Forschungsgemeinschaft (DFG) grant SCHN 520/8-1.
R.C.G.\ gratefully acknowledges the hospitality
of the Grupo de F\'{\i}sica No Lineal (GFNL, University of Sevilla, Spain)
and support from NSF-DMS-0806762, Plan Propio de la Universidad de Sevilla,
Grant No. IAC09-I-4669 of Junta de Andalucia and Ministerio de Ciencia e
Innovaci\'on, Spain.
P.G.K.\ acknowledges the support from NSF-DMS-0806762, NSF-CMMI-1000337,
as well as from the Alexander von Humboldt Foundation and from the 
Alexander S. Onassis Public Benefit Foundation.


\end{document}